\documentclass[12pt]{article}
\usepackage{amsmath,amssymb}
\usepackage{epsfig,rotating}

\newcommand{\as}{\alpha_s}

\newcommand{\mug}{\mu_G^2}

\newcommand{\ep}{\epsilon }
\newcommand{\uh}{{\hat u} }
\newcommand{\zh}{{\hat z} }

\long\def\symbolfootnote[#1]#2{\begingroup%
\def\thefootnote{\fnsymbol{footnote}}\footnote[#1]{#2}\endgroup}

\def \be{\begin{equation}}
\def \ee{\end{equation}}
\newcommand{\bea}{\begin{eqnarray}}
\newcommand{\eea}{\end{eqnarray}}
\def \nn{\nonumber}

\numberwithin{equation}{section}
\allowdisplaybreaks[4]

\begin{document}
\begin{titlepage}
\flushright{TTP12-041\\
SFB/CPP-12-83\\}
\vspace*{3.5cm}

\centerline{\Large\bf\boldmath Kinetic operator effects in $\bar B\to X_c\ell\nu$ at $O(\alpha_s)$}

\vskip 2.5cm

\begin{center}
  {\bf Andrea Alberti$^{a}$, Thorsten Ewerth$^b$, \\[2mm]
  Paolo Gambino$^a$, Soumitra
    Nandi$^{c,d}$}\\[5mm]
  {\sl $^a$ Universit\`a di Torino, Dip.\ di  Fisica \& INFN Torino, I-10125, Italy}\\[2mm]
    {\sl $^b$ Institut f{\"u}r Theoretische Teilchenphysik,\\
Karlsruhe Institute of Technology (KIT),
D-76128, Germany}\\[2mm]
    {\sl $^c$ Theor.\  Physik 1, Fachbereich Physik, Univ.\ Siegen, D-57068, Germany}\\[2mm]
       {\sl $^d$ Dept.\ of Physics,
Indian Inst.\ of Technology Guwahati, 781 039, India}
\end{center}

\vskip 3cm

\begin{abstract}
  We compute the $O(\alpha_s)$ corrections to the Wilson coefficient of the kinetic operator 
  in inclusive semileptonic $B$ decays. 
  Our analytic calculation agrees with reparameterization invariance and with  previous numerical results and paves the way 
  to the calculation of analogous corrections to other power-suppressed coefficients.
  
  \end{abstract}

\end{titlepage}

\section{Introduction}
The precision determination of the Cabibbo-Kobayashi-Maskawa quark mixing 
 matrix remains a central goal in the flavor physics 
program. A  new generation of high-luminosity $B$ factories is expected to start
operations in a few years and in view of the improved experimental resolution
the theoretical uncertainties should be reduced whenever 
possible.  In the case of inclusive semileptonic $B$ decays, which currently provide the
most precise determination of $|V_{cb}|$ and $|V_{ub}|$, theoretical uncertainties are already dominant, but there is space for improvement. 

As is well known, the theoretical foundation for  our understanding of 
 inclusive semileptonic 
decays $B\to X_c \ell\nu$
is  an Operator Product Expansion (OPE) which ensures that  non-perturbative 
effects are suppressed by at least two powers of the bottom mass $m_b$. They are  
parameterized by a limited number 
of matrix elements of local operators which  can be extracted from experimental data.
The total inclusive width and the first few moments of the  kinematic distributions can 
 be well approximated by a
double series in $\as$ and $\Lambda_{\rm QCD}/m_b$ \cite{Bigi:1992su,Blok:1993va}.
After extracting the 
most important non-perturbative parameters, including the heavy quark masses, from the 
moments, one can therefore use them in the OPE expression for the total semileptonic  width and  determine  $|V_{cb}|$ from the comparison with the experimental  rate.

It is worth emphasizing that the information obtained from the fits to the moments  
of $B\to X_c \ell\nu$ \cite{fits,ckm10,hfag}
find other important applications. Indeed, the
$b$ quark mass and the OPE expectation values obtained from the moments 
are crucial inputs in the determination of $|V_{ub}|$ from inclusive semileptonic charmless decays, see {\it e.g.}\ \cite{Antonelli:2009ws} and refs.\ therein. 
The heavy quark masses and the OPE parameters 
are also relevant in the calculation of inclusive rare decays like $B\to X_s 
\gamma$  (see \cite{bsgamma} for the state of the art calculation and \cite{review} for a review), in particular for the normalization of their branching ratios \cite{gg}.

The reliability of the inclusive method rests on our ability to control 
the higher order contributions in the OPE. If we neglect perturbative corrections, {\it i.e.}\ if we work at tree-level,  
we presently know the $O(1/m_b^2)$ and $O(1/m_b^3)$ contributions \cite{1mb3}, while the 
 $O(1/m_b^4)$ and $O(1/m_Q^5)$ effects have been studied in \cite{Mannel:2010wj}.
 Unfortunately, new non-perturbative parameters appear at each order in the OPE: 
 as many as nine new expectation values appear at $O(1/m_b^4)$. As a result, only the parameters associated with the $O(1/m_b^{2,3})$ corrections are routinely fitted from experiment. 
In \cite{Mannel:2010wj} the parameters associated with $O(1/m_Q^{4,5})$ effects 
 have been estimated  in the ground state saturation approximation,  
finding a relatively small +0.4\% net effect on $|V_{cb}|$. Recently, the validity of
the ground state saturation has been investigated \cite{Gambino:2012rd}, and it has been 
shown that  the non-factorizable contributions can be in general comparable to the 
factorizable ones.   Additional work is therefore necessary to assess the importance of 
higher order power effects.

For what concerns the purely perturbative corrections to the free quark decay, they are known at $O(\as^2)$ in all the relevant cases, namely for  the width and the first few moments  of  the lepton energy and hadronic mass distributions. 
The  complete $O(\as)$ and $O(\alpha_s^2 \beta_0)$ corrections have been computed some time ago \cite{pert,Aquila}, while  the remaining two-loop  corrections to the width and to 
the first few moments have been calculated  in Refs.~\cite{melnikov,czarnecki-pak,melnikov2,Gambino:2011cq}. The theoretical uncertainty due to missing 
purely perturbative effects is now relatively small \cite{Gambino:2011cq}.

The  $O(\as \Lambda^2_{\rm QCD}/
m_b^2)$ corrections  appear to  be a potentially more important source of theoretical uncertainty. The $O(\as)$ corrections to the 
Wilson coefficient of the kinetic operator have been computed  numerically in \cite{Becher:2007tk}; they  can be also obtained from the 
parton level $O(\as)$ result using reparameterization invariance  relations 
\cite{Bigi:1992su,RPI,Manohar:2000dt,Manohar:2010sf}. 
They lead to numerically modest $O(\as \mu^2_{\pi}/m_b^2)$ corrections to the width and moments, where $ \mu^2_{\pi}$ is the matrix element of the kinetic operator. 
However, in order to have all the  $O(\as \Lambda^2_{\rm QCD}/
m_b^2)$ effects one should also consider the  $O(\as)$ corrections to the Wilson coefficient 
of the chromomagnetic operator.  
A complete $O(\as \Lambda^2_{\rm QCD}/
m_b^2)$ calculation  has been performed 
in the simpler case of inclusive radiative decays \cite{ewerth}, where the $O(\as)$ corrections increase the 
coefficient of $\mug$, the matrix element of the chromomagnetic operator, by almost 20\% in the rate. 

In this paper we present the first part of an analytic calculation of the 
$O(\as \Lambda^2_{\rm QCD}/m_b^2)$ corrections. We extend the method developed in \cite{ewerth} to 
semileptonic decays into hadronic final states containing a massive quark and validate it rederiving the  $O(\as \mu^2_{\pi}/m_b^2)$ 
corrections and reproducing the reparameterization relations. 
As we did in \cite{ewerth}, we compute the relevant Wilson coefficients at $O(\alpha_s)$ by 
Taylor expanding off-shell amputated Green functions  around the $b$ quark mass shell, and by 
matching them onto local operators in Heavy Quark Effective Theory (HQET). 
The extension to semileptonic decays implies new technical difficulties, because  one needs to consider the mass of the final quark 
and the non-vanishing invariant mass of the 
lepton pair. The integrals involved are therefore less divergent but more complex.
We identify a small number of master integrals and  express our results in terms 
of the same functions appearing in the $O(\as)$ parton calculation. In order to 
compute contributions to arbitrary moments, we give explicit corrections to the three 
independent   structure functions, namely of the triple differential rate.
The 
$O(\as \mu^2_{G}/m_b^2)$ calculation is under way and will be presented in a forthcoming publication.

The outline of this paper is as follows. In section 2 we introduce our notation and review the 
known $O(1/m_b^2)$  and $O(\as)$ corrections to the triple differential rate.
Section 3 is devoted to a description of the calculation of the $O(\as \mu_\pi^2/m_b^2)$ contributions.
Our analytic results can be found in section 4.  In section 5 we review the reparameterization invariance constraints and show that they are satisfied by our results. In section 6
we summarize and
conclude. The relevant master integrals and a few technical details are given in the Appendix.

\section{Leading Order results}
We start recalling the tree-level results and the $O(\as)$ corrections to the leading, free-quark  term of the OPE. We consider the decay of a $B$ meson 
of four-momentum $p_B= M_B v$ into a lepton pair with momentum $q$ and 
a hadronic final state containing  a charm quark with momentum $p'=p_B-q$.
The hadronic tensor $W^{\mu\nu}$ which determines the hadronic contribution to the differential 
width  is given by the absorptive part of a current correlator in the appropriate kinematic region,
\be
W^{\mu\nu}(p_B,q)= {\rm Im} \,  \frac{2\,i}{\pi M_B}  \int d^4 x\, e^{-i q\cdot x} \langle \bar B| T J_L^{\mu\dagger}(x) J_L^{\nu} (0) | \bar B\rangle,
\label{correlator}
\ee
where $J_L^\mu=\bar c \gamma^\mu P_L b$ is the charged weak current. The correlator is
subject to an OPE in terms of local operators, which at the level of the 
differential rate takes the form of an 
expansion in inverse powers  of the energy  release, whose leading term corresponds to the 
decay of a free quark. 

We generally follow the notation of Ref.~\cite{Aquila} and express the $b$-quark decay kinematics in terms of the dimensionless quantities 
\be
\rho= \frac{m_c^2}{m_b^2}, \qquad\quad \hat u= \frac{(p-q)^2 - m_c^2}{m_b^2} ,\quad\qquad \hat q^2= \frac{q^2}{m_b^2},
\ee
where $p= m_b v$ is the momentum of the $b$ quark and 
\be
0\le \hat u \le \hat u_+ =(1-  \sqrt{\hat{q}^2})^2 -\rho \qquad  {\rm and} \qquad 
0\le \hat q^2 \le (1-\sqrt{\rho})^2.
\label{physrange}
\ee
We will also employ the energy of the hadronic system normalized to the $b$ mass 
\be
E=  \frac12 (1+\rho +\hat u -\hat q^2).
\ee
The case of tree-level kinematics corresponds to $\hat u=0$; we indicate the corresponding   energy of the hadronic final state as
\be 
E_0= \frac12 (1+\rho -\hat q^2).
\ee
 The  normalized total leptonic energy is
\be
\hat q_0 =1-E \quad\quad {\rm from\ which\ follows } \quad\quad \hat u = 2\,(1-E_0-\hat q_0).
\ee
We also introduce a threshold factor 
\be
\lambda=4\,(\hat q_0^2-\hat q^2)=
4\,(E^2-\rho-\hat u). 
\ee
In the case of tree-level kinematics, the threshold factor becomes $\lambda_0=4(E_0^2-\rho)$. 
It is convenient to introduce a short-hand notation for the square root of $\lambda$:
\be
t= \frac{\sqrt{\lambda}}{2\, E}, \qquad\quad t_0= \frac{\sqrt{\lambda_0}}{2\, E_0}.
\ee

It is customary to decompose the hadronic tensor
as follows
\be
m_b \,W^{\mu\nu}(p_B,q)=-W_1 \, g^{\mu\nu}+W_2 \,v^\mu v^\nu +i W_3 \, \ep^{\mu\nu\rho\sigma }
v_\rho \hat q_\sigma + W_4 \hat q^\mu \hat q^\nu +W_5 \left(v^\mu \hat q^\nu\!+\! v^\nu \hat q^\mu\right),
\ee
where  the structure functions $W_i$ are functions of $\hat q^2, \hat q_0$ or equivalently of
$\hat q^2, \hat u$, $v^\mu$ is the four-velocity of the $B$ meson, and $\hat q^\mu=q^\mu/m_b$. 

 In the limit of massless leptons only $W_{1,2,3}$ contribute to the decay rate and one has
\bea
\frac{d \Gamma}{\, d\hat E_\ell \,d\hat q^2 \, d\hat u }&=&
\frac{G_F^2 m_b^5 |V_{cb}|^2}{16\pi^3}
\theta(\hat u_+ -\hat u)\times \\ &&
\times\left\{
\hat q^2\, W_1 -\left[2 \hat E_\ell^2-2\hat E_\ell\hat q_0 +\frac{\hat q^2}2 \right] W_2 +\hat q^2 (2\hat E_\ell -\hat q_0)\,W_3\right\} ,\nn
\label{rate}
\eea
where $\hat u_+= (1-\sqrt{\hat q^2})^2-\rho$
represents the kinematic boundary on $\hat u $, and $\hat E_\ell=E_\ell/m_b$ is the normalized charged lepton energy.
Thanks to the OPE, the structure functions can be expanded in series of $\alpha_s$ and $\Lambda_{\rm QCD}/m_b$. There is no term linear in $\Lambda_{\rm QCD}/m_b$ and therefore
\be
W_i =W_i^{(0)} + \frac{\mu_\pi^2}{2m_b^2} W_i^{(\pi,0)}+\frac{\mu_G^2}{2m_b^2} W_i^{(G,0)}+
  \frac{C_F\as}{\pi}\left[  W_i^{(1)} +
  \frac{\mu_\pi^2}{2m_b^2} W_i^{(\pi,1)}+\frac{\mu_G^2}{2m_b^2} W_i^{(G,1)}\right]
\ee
where we have neglected terms of higher order in the expansion parameters.
$\mu_\pi^2$ and $\mu_G^2$ are the $B$-meson matrix elements of the only gauge-invariant 
dimension 5 operators that can be formed from the $b$ quark and gluon fields  \cite{Bigi:1992su,Blok:1993va}.
The leading order coefficients are given by 
\be
W_i^{(0)} = w_i^{(0)} \,\delta(\hat u);  \qquad  \qquad w_1^{(0)} = 2 E_0, \qquad w_2^{(0)} = 4, \qquad w_3^{(0)} =2.
\ee
The tree-level nonperturbative coefficients  \cite{Blok:1993va}  read
\be
W_i^{(\pi,0)} = w_i^{(\pi,0)} \,\delta(\hat u) +w_i^{(\pi,1)} \,\delta'(\hat u) +
 w_i^{(\pi,2)} \,\delta''(\hat u); 
 \ee
   
 \begin{center}
 \begin{tabular}{lll}
$w_1^{(\pi,0)}= \frac83(1- E_0),$ \quad& $ w_1^{(\pi,1)}= \frac43 E_0(1-E_0),$\quad & \quad$w_1^{(\pi,2)}=\frac{2}3E_0 \lambda_0;$\nn\\[1mm]
$w_2^{(\pi,0)}= 0, $\quad&$ w_2^{(\pi,1)}= -8(1-E_0), $\quad&\quad$ w_2^{(\pi,2)}=\frac{4}3 \lambda_0 ;$\nn\\[1mm]
$ w_3^{(\pi,0)}= -2,$ \quad&$ w_3^{(\pi,1)}= -\frac43 (1-E_0), $\quad&\quad$ w_3^{(\pi,2)}=\frac{2}3\lambda_0,$\nn
\end{tabular}
\end{center}
and 
\be
W_i^{(G,0)} = w_i^{(G,0)} \,\delta(\hat u) +w_i^{(G,1)} \,\delta'(\hat u);  
\ee
 \begin{center}
 \begin{tabular}{ll}
$w_1^{(G,0)}= -\frac43 (2-5E_0),$ \quad& $ w_1^{(G,1)}= -\frac43 (E_0+3E_0^2+\frac12 \lambda_0);$ \nn\\[1mm]
$w_2^{(G,0)}= 0, $\quad&$ w_2^{(G,1)}=\frac83 (3-5E_0)   ; $ \nn\\[1mm]
$ w_3^{(G,0)}=\frac{10}3 ,$ \quad&$ w_3^{(G,1)}= -\frac43 (1+5E_0).$\nn
\end{tabular}
\end{center}

The perturbative corrections to the free quark decay have been computed in \cite{Aquila} and refs.\ therein. They read
\be
W^{(1)}_i=w_i^{(0)} \left\{ S_i \,\delta(\hat u) -2 \left(1 -  E_0 I_1\right)
 \left[\frac1{\hat u}\right]_+   +\frac{\theta(\hat u)}{(\rho + \hat u)}   \right\} + R_i  \,\theta(\hat u)
\label{NLO},
\ee
where $S_i= S+\Delta_i$ and
\bea
S&=&2E_0 \left(I_{2,0}-I_{4,0}\right) 
-1 -\frac{1-\rho-6\hat q^2}{4\hat q^2} \ln \rho 
-\frac{(1 - \rho)^2 - 6\,  \hat q^2 (1 + \rho) + 5 (\hat q^2)^2 }{4\hat q^2}  I_{1,0}\,;\nn\\
 \Delta_1\!&=&\!\!-\frac{\rho}{E_0} I_{1,0}; \qquad \Delta_2= \frac{1-\rho}{4\hat q^2} \ln \rho+\left(\frac{(1-\rho)^2}{4\hat q^2}-\frac{1+\rho}{4}\right)I_{1,0}; \qquad \Delta_3=0, \nn
\eea
and the functions $R_i$ are given in Eqs.~(2.32-2.34) of Ref.~\cite{Aquila}.\footnote{The
variables $\hat \omega $, $\lambda_b$, and $\tau$ of Ref.~\cite{Aquila} correspond to $-2 E_0$, $\lambda$ and $(1-t)/(1+t)$, respectively.}
 The integrals 
$I_1$, $I_{1,0}$, $I_{2,0}$, and $I_{4,0}$ are given below in Eqs.~(\ref{I1},\ref{I40},\ref{I20})
and the plus distribution is defined by
\be
\left[\frac1{\hat u}\right]_+ =\lim_{\varepsilon\to 0} \left[ 
\ln \varepsilon \, \delta(\hat u) + \frac1{\hat u} \,\theta(\hat u - \varepsilon)  \theta(1-\hat u)
\right]
\label{plus1b}
\ee
or equivalently by its action on a test function $f(\hat u)$:
\be
\int d\hat u \,f(\hat u) \left[\frac1{\hat u}\right]_{+} = 
\int_0^{1} d\hat u \ \frac{f(\hat u)-f(0)}{\hat u} .
\label{plus1}
\ee
The upper  limit of integration in the rhs of (\ref{plus1}) can be chosen arbitrarily, but it is convenient to have it larger than the physical boundary, $\hat u_+$.  
Ref.~\cite{Aquila} uses $\hat u_+$ as upper limit, and the two definitions are  related by the simple expression
\be
\left[\frac1{\hat u}\right]_{+,[10]} =\left[\frac1{\hat u}\right]_+ -
\ln \hat u_+ \,\delta(\hat u)-
\frac{\theta(\hat u - \hat u_+)\,\theta(1-\hat u)}{\hat u}.
\ee

\begin{figure}[t]
\begin{center}
\includegraphics[width=12cm]{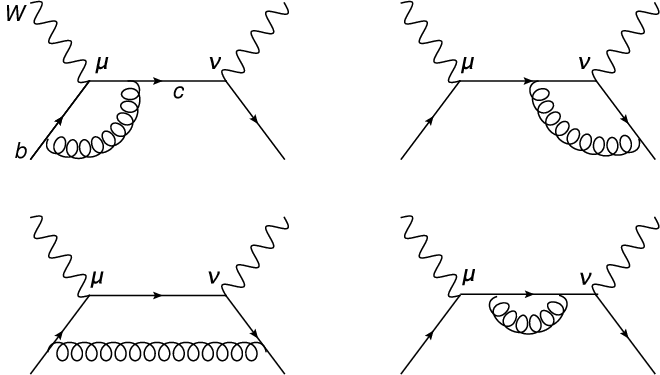}
\caption{\sf One-loop diagrams contributing to the current correlator.}
\label{fig1}
\end{center}
\end{figure}

\section{The calculation of $O(\as\mu_\pi^2/m_b^2)$ effects}
The four diagrams in Fig.~1 are our starting point. They contribute to 
 the weak current correlator of Eq.~(\ref{correlator}) and are sufficient to 
determine both $W^{(1)}_i$ and  $W_i^{(\pi,1)}$, while additional diagrams with external background gluons are 
necessary for the determination of $W_i^{(G,1)}$.  The  momenta of the external 
$b$ quarks and $W$ bosons are $p=m_b v +k$ and $q$, respectively. 
We write down the corresponding off-shell
forward amplitudes, and extract the contributions to $W_{1,2,3}$ by contraction with 
appropriate tensor projectors. We then 
Taylor expand around $k=0$, {\it i.e.} around the mass-shell of the $b$ quarks, 
through $O(k^2)$.  We always work 
in $d=4-2\epsilon$ dimensions and use dimensional regularization for both ultraviolet and 
infrared divergences. The result of the Taylor expansion 
is reduced to  scalar integrals, which are in turn expressed in terms of 4 independent master 
integrals, listed in the Appendix, using Integration by Parts (IBP) identities \cite{IBP}.

The forward amplitudes obtained in this way correspond to quark matrix
elements of local operators which eventually 
have to be evaluated in the $B$ meson. In particular,
the $k$-dependent coefficients of the master integrals should be expressed in terms of 
$B$ meson matrix elements of local operators. To this end it is convenient to use HQET: all 
$O(k)$ and $O(k^2)$  matrix elements can be expressed in terms of 
\begin{equation}\label{eq::lambdai}
  \lambda_1 = \frac{1}{2m_B}\langle\bar B(v)|\bar b_v(iD)^2b_v|\bar B(v)\rangle\,,\qquad
  \lambda_2 = -\frac{1}{6m_B}\langle\bar B(v)|\bar b_v\frac{g_s}{2}G_{\mu\nu}\sigma^{\mu\nu}b_v|\bar B(v)\rangle\,,
\end{equation}
where  $D_\mu=\partial_\mu+ig_sG_\mu^aT^a$ is the 
covariant derivative, $G_{\mu\nu}$ the gluon field tensor,  and $b_v$ is the static quark field (see \cite{ewerth} and refs.\ therein for details).
While $\lambda_1$ and $\lambda_2$ are defined in the asymptotic HQET regime, in practical applications one deals
with $\mu_\pi^2= -\lambda_1+O(1/m_b)$ and $\mu_G^2=3\lambda_2+O(1/m_b)$, defined  
in terms of full QCD states at finite $m_b$. The power corrections to these relations are irrelevant 
for our calculation. In this paper we consider only the terms proportional to $\lambda_1$.

The next step to compute the physical structure functions $W_i$ consists in taking the 
imaginary part of the result.  There are two kinds of contributions to the imaginary part: 
the first  comes from the imaginary part of a charm propagator, raised to power $n$, outside of the loop,  via 
\be
\frac1{\pi}\,{\rm Im}\, \frac{1 
}{\big[ (v-\hat q)^2 -\rho^2 +i \eta\big]^n} = \frac{(-1)^n}{n!}\delta^{(n-1)}(\hat u) ;
\ee
the other comes from the imaginary part of a loop integral, and is related to real gluon 
emission. In the first case the real part of the loop integrals is multiplied by 
$\delta(\hat u)$ or its derivatives. It follows that the real parts of the  
master integrals, together with their derivatives wrt $\hat u$, are only needed at $\hat u=0$,
{\it i.e.} with partonic kinematics. The derivatives
of the master integrals wrt $\hat u$ can be, in turn, re-expressed in terms of the 
same master integrals, for example using
\be
\frac{d}{d\hat u}\Big |_{{\rm fixed} \ q^2}= \frac{\partial q_0}{\partial \hat u}\frac{\partial q_{\mu}}{\partial  q_0}\frac{\partial}{\partial q_{\mu}}\Big |_{{\rm fixed}\ q^2}=-\frac{1}{2}  \frac{(\hat q^2 p_{\mu}- \hat q_0 q_{\mu} )}{( \hat q^2- \hat q_0^2)}\frac{\partial}{\partial q_{\mu}}
\label{deriv}.
\ee
As a result, the contribution to the final result coming from the real parts of the loop integrals
can be expressed in terms of a single combination of dilogarithms and a single logarithm,
defined as $I_{4,0}$ and $I_{1,0}$ in Eqs.~(\ref{I1},\ref{I40}). The latter are   
the same functions that appear in Eq.~(\ref{NLO}). 

All the singularities are located at the threshold, $\hat u=0$. We therefore 
identify in the master integrals 
the terms which potentially lead to infrared divergences and
employ the identity 
\be
\uh^{-A+B\ep}= \sum_{p=0}^{A-1} \frac{(-1)^p}{p!} \frac{\delta^{(p)}(\uh)   
 }{1+p-A+B\ep} +\sum_{n=0}^{\infty}\frac{(B\ep)^n}{n!} \left[\frac{\ln^n \uh}{\uh^A}\right]_+ \,,
\ee
valid for $A>0$, where the plus distributions are defined by generalizing Eq.~(\ref{plus1}),
\be
\label{plusn}
\int d\uh \left[\frac{\ln^n \uh}{\uh^m}\right]_+  f(u)=\int^{1}_0 d\uh\,
\frac{\ln^n \uh}{u^m} \left[f(u)-\sum_{p=0}^{m-1} \frac{u^p}{p!} f^{(p)} (0)\right]
\ee
with $f^{(p)}(u)= \frac{d^p f(u)}{d u^p}$. These terms enter the imaginary part of the master 
integrals and control the infrared divergences due to real gluon emission.

In the calculation  of $W_{1,2}$ the problem of the $d$-dimensional definition of $\gamma_5$ can
be avoided by simply anticommuting it with all $\gamma^\mu$ matrices.
In the case of the parity
violating structure function $W_3$, however, one needs to proceed with care and  
adopt a $d$-dimensional definition for the axial-vector current. One possibility is to follow  
\cite{larin}  and employ the replacement
\be 
\bar c \gamma^\mu \gamma_5 b \to -i \,\frac1{3!} \epsilon^{\mu\nu\rho\sigma} \,\bar c \gamma_\nu\gamma_\rho \gamma_\sigma b,
\ee
where $\epsilon^{\mu\nu\rho\sigma}$ is a strictly 4-dimensional object. 
This Levi-Civita tensor is multiplied by another antisymmetric $\epsilon$ tensor,  
necessary to extract the $W_3$ component of the amplitude, and their product 
can be expressed as a combination of metric tensors. The latter can then be 
taken in $4-2\epsilon$ dimensions. 
This definition has several advantages but it violates chiral symmetry and its basic Ward identities. Therefore it requires the introduction of  a finite one-loop renormalization constant $Z_5$
\be 
\bar c \gamma^\mu \gamma_5 b \to -i \,\frac1{3!} \epsilon^{\mu\nu\rho\sigma} \,Z_5\,\bar c \gamma_\nu\gamma_\rho \gamma_\sigma b,
\ee
of course in addition to the wave-function renormalization of the external legs. $Z_5$ is given in \cite{larin} and Refs. therein for the massless case, 
\be
Z_5= 1- C_F\, \frac{\as}{\pi},
\ee
and we have checked that this result applies to the case of massive quarks as well.
Another possibility, which leads to the same result without an extra finite renormalization, is to 
anticommute $\gamma_5$ to the extreme left of the Dirac string in all diagrams and then to replace it by its four-dimensional definition. 

After combining all the diagrams, the infrared divergences are cancelled and the ultraviolet 
divergences are removed by the $b$ quark wave function renormalization and the 
charm mass renormalization. The charm mass renormalization also removes all the $\delta'''(\hat u)$ terms. No renormalization is necessary in the effective theory \cite{ewerth}.

\section{The  $O(\as \mu_\pi^2/m_b^2)$ results}
We now report our  results for the $O(\as)$ corrections to the Wilson coefficients of the kinetic operator. The most singular part of the $W_i$s has a universal structure 
\be
B_{(i,\pi)}= \frac{\lambda_0}3   \Big\{\Big[  S_i +3(1-E_0 I_{1,0})\Big] \delta''(\hat u) 
-  4 \left(1 -  E_0 I_{1,0}\right)
 \left[\frac1{\hat u^3}\right]_+ \Big\},
\ee
and the complete results are 
\bea
W^{(\pi,1)}_1&=&
      2E_0 \,B_{(1,\pi)}+\frac83(1-E_0 I_{1,0} )(1-E_0)
\left(     
 E_0\left[\frac1{\hat u^2
 }\right]_+  -2 \left[\frac1{\hat u
 }\right]_+ \right)   + R_1^{(\pi)}  \,\theta(\hat u)  \nn\\
  &&
\hskip-1cm+\frac{8E_0}3\left[ \frac{1-E_0}2  S_1- \lambda_0(1-E_0 I_{1,0}) 
-\frac{E_0^2}{\rho}\! +
\!       (  3 E_0^2\! -\! \rho) I_{1,0} - 2 E_0 (1\! +\! I_{1,0}) + 3   \right]\delta'(\hat u)   \nn\\
&&\hskip-1cm+\frac83
\left[ S_2
  -E_0 S_1  +E_0 (1\!-\!E_0 I_{1,0})\left(\frac{\lambda_0}2 -
 \frac{(1-E_0)^2}{\lambda_0} +E_0\right)+ \left( E_0^2 - \frac34 E_0 - \frac{\rho}4\right) I_{1,0}
\right. \nn \\
&& \hskip-1cm\left.
 + \frac{\frac34 E_0^3 -     \frac{\rho}4 (E_0^2 + 2 E_0 - 2 \rho (1 - E_0 )) }{
  \rho^2} 
 \right]\delta(\hat u), 
\eea
\bea
W^{(\pi,1)}_2&=&
      4 B_{(2,\pi)} -16 (1-E_0 I_{1,0} )
             (1-E_0)\left[\frac1{\hat u^2
 }\right]_+      + R_2^{(\pi)}  \,\theta(\hat u)  \\
  &&\hskip-1.2cm -8\left[ (1\!-\!E_0) S_2 + 2- \frac{8}3 E_0
  +\frac{2}3 \lambda_0 (1\!-\!E_0 I_{1,0}) \!    +\!\Big(\frac{\lambda_0}2 
   + \frac{8}3  \rho-\frac83E_0\Big)  I_{1,0}
  +   \frac{2 E_0^2}{3\rho}    
    \right]\delta'(\hat u)   \nn\\&&\hskip-1.2cm-8
\Big[ 
  (1\!-\!E_0 I_{1,0}) \Big(2 E_0 \!-\!\frac{\lambda_0}3 \!-\!7
 \frac{(1\!-\!E_0)^2}{\lambda_0} \Big)
 - \frac{     E_0^2 -\rho E_0 + \frac{\rho}6  }{
  2\rho^2} -\frac{5}3+ \left(\frac23+\frac{13E_0}{12}  \right) I_{1,0}
 \Big]\delta(\hat u) , \nn   
\eea
\bea
W^{(\pi,1)}_3&=&
      2 B_{(3,\pi)}-4(1- E_0 I_{1,0} )\left(
\frac23(1-E_0) \left[\frac1{\hat u^2
 }\right]_+   -\left[\frac1{\hat u
 }\right]_+ \right)  + R_3^{(\pi)}  \,\theta(\hat u) \nn\\
  &&\hskip-1.2cm- \frac43\left[(1-E_0) S
  +2\lambda_0 (1-E_0 I_{1,0})   -2 \Big( I_{1,0} -\frac{E_0}{\rho}\Big)(E_0 - \rho)
    \right]\delta'(\hat u)  \\
&& \hskip-1.2cm-2  \left[S-\Big(\frac{4
   (1\!-\!{E_0})^2}{{\lambda_0}}-\frac{4 {E_0}}{3}+\frac{2
   {\lambda_0}}{3}+1\Big) (1\!-\!{E_0} {I_{1,0}}) +(1\!-\!{E_0}) {I_{1,0}}-\frac{{E_0}
   ({E_0}\!-\!{\rho})}{{\rho}^2}\right] \delta(\hat u).
 \nn 
 \eea
In the above expressions the coefficients of the derivatives of $\delta(\hat u)$ have been 
reduced using integration by parts: for instance
\be
f(\hat u) \,\delta '' (\hat u)= f(0)\, \delta''(\hat u) -2 f'(0)\, \delta'(\hat u) + f''(0)\, \delta(\hat u).
\ee
 The coefficients of the plus distributions can be similarly reduced by Taylor expanding them around $\hat u=0$, for example:
 \be
 f(\hat u) \left[\frac1{\hat u^2}\right]_+=f(0)\left[\frac1{\hat u^2}\right]_+ + f'(0) \left[\frac1{\hat u}\right]_+ + \frac{f(\hat u)-f(0)-\hat u f'(0) }{\hat u^2} ,
 \ee
where the last term is regular for $\hat u\to 0$.

The functions $R_i^{(\pi)}$ are related to real gluon emission and are given by 
\bea
R_1^{(\pi)}&=& \frac83E_0\Big( \frac{(1\!-\!E_0)(E_0-2\hat u)}{\hat u^2}  -\frac{E_0 \lambda_0}{\hat u^3}\Big)I_{1,0} -
\frac{\frac43 { \hat u^2}+2   (1\!-\!5E) \hat u\!+\!\frac{28}3
   (E^2\!-\!E)}{\lambda }\Big(I_1-\frac1{E}\Big)
\nn\\&&\!\!\!+
\frac83\left(\frac78 E  -\frac38 - \frac{\hat u}4 
+ \frac{E - z}{\hat u}+ \frac{E^2 \lambda}{\hat u^3} + \frac{E (E - E^2 - \lambda)}{\hat u^2} \right) I_1  +\frac{16 E^2 (E-z)}{3 \hat u^2 z}\nn\\
  &&\!\!\!
 - \frac{\hat u^2 \left(4 E^2-E z-2 z^2\right)}{6
   E z^3}
   +\frac{-8 E^3-3 E^2 z+5 E z (z+1)+(5-2 z) z^2}{3 z^3}
   \\&&\!\!\!
   + \frac{4 E (2 E^2-5 E z+4 z^2  )}{3\hat u\, z^2}+
   \frac{\hat u \left(24 E^4-12
   E^3 z+4 E^2 (z-3) z+E \left(z^2-6 z^3\right)+3
   z^3\right)}{6 E z^4},\nn
\eea
\bea
R_2^{(\pi)}&=&  -\frac{16}3
\frac{ {E_0} {\lambda_0} }{{\hat u }^3}(I_{1,0}\!-\!I_1)-\frac{16 (1\!-\! E_0) {E_0}
  }{{\hat u}^2}  \Big({I_{1,0}}-\frac{{I_1}}{3}\Big) -\frac23 \Big[16E_0\frac{
    4 (1\!-\!{E_0})^2 -\frac{
   {\lambda_0}}{4}}{{\lambda_0} {\hat u}}\nn\\
 &&  -64 \frac{ (1\!-\! {E_0})^2 E_0}{{\hat u}} \Big(\frac{1}{{\lambda_0}}-\frac{1}{\lambda }\Big)
 +\frac{70 {E}^2-{E} (71 {\hat u}+94)+10
   {\hat u}^2+43 {\hat u}+24}{ \lambda }  +2\Big]I_1\nn\\
   &&+\frac{8
   \big({E}\!-\!1)^2 (14 ({E}^2-E) -{\hat u} (15 {E}-2
   {\hat u}-3)\big)}{\lambda ^2}  \Big( I_1-\frac1{E}\Big)
   +\frac{8 {E_0} {E} {\hat u}}{z^4}\nn\\
   &&
   +\frac{2 \big(6 {E} {\hat u}-8E^2+{\hat u} (3 {\hat u}\!-\!4)\big)}{3 z^3} 
   +\frac{\frac{8 \left(2
   {E}^2-7 {E} z+2 z (z+2)\right)}{3 z^2}+\frac{128
   ({E}-1)^2}{3 \lambda }}{{\hat u}}  -\frac{16 (5
   {E}+{\hat u}+4)}{3 \lambda }\nn\\
   &&
   +\left(\frac{(2 {E}-1) {\hat u}^2}{3 {E}^3}+\frac{54
   {E}^2-25 {E}+7}{3 {E}^2}-\frac{\left(22
   {E}^2-21 {E}+3\right) {\hat u}}{6 {E}^3}\right)
   \left(\frac{4}{\lambda }+\frac{1}{z}\right)\nn\\
   &&
   +\frac{\frac{\frac{2
   {\hat u}^2}{3}+{\hat u}}{{E}}-10 {E}-7
   {\hat u}+10}{z^2}+\frac{32 {E} ({E}-z)}{3 {\hat u}^2 z},
\eea
\bea R_3^{(\pi)}&=& \! \!\!-\frac{8 {E_0}
   \left(\lambda_0-{E_0} {\hat u}+\left(1-\frac{3
   {\hat u}}{2}\right) {\hat u}\right)}{3 {\hat u}^3}  I_{1,0}
   +
   \left(\frac{8 (3 {E}-5 {E}^2+2 z)}{3
   {\hat u}^2}+\frac{8 {E} \lambda }{3
   {\hat u}^3}-\frac{4-\frac{8
   {E}}{3}}{{\hat u}}-\frac{2}{3}\right)
I_1\nn\\
&&   
  - \frac{4 (1\!-\!{E}) \big(8 (E\!-\!{E^2})
   -{\hat u} (1\!-\!9 {E}\!+\!2 {\hat u})\big)}{3 \lambda \,{\hat u}} \Big({I_1}-\frac{1}{{E}}\Big)
   +\frac{4 {E}^2
   {\hat u}}{z^4}-\frac{2 \left(4 {E}^2+3 {E}
   {\hat u}+{\hat u}\right)}{3 z^3}
  \nn\\&& 
   +
   \frac{8 {E}^2-20 {E} z+8 z (z+1)}{3 {\hat u} z^2}
   +\frac{16 {E} ({E}-z)}{3 {\hat u}^2 z}+\frac{7-3
   {E}-2 {\hat u}}{3 z^2}+\frac{7 {E}-2 {\hat u}-1}{3
   {E} z},
\eea
 with $z= \hat u +\rho$.  Of course, the functions $R_i^{(\pi)}$ are regular in the limit $\hat u \to 0$. 

The above results can be used in Eq.~(\ref{rate}) to compute the $O(\as \mu_\pi^2/m_b^2)$
corrections to the total rate and to the moments of various differential distributions. 
The phase space integration is rather delicate because of strong cancelations between different 
singular terms, especially in the presence of a cut on the lepton energy. 
We have compared the numerical results with those in Tables 1 and 2 of Ref.~\cite{Becher:2007tk} and found agreement in all cases.
In principle, it is also possible to take the limit $\rho\to 0$ and obtain analytic results for the $B\to X_u \ell\nu$ decay.

\section{Reparameterization Invariance relations}
Reparameterization Invariance (RI) \cite{RPI,Manohar:2000dt}
connects different orders in the heavy quark expansion.
In particular, as we have mentioned in the Introduction, 
RI links the coefficient of the kinetic operator to the coefficient of the leading, dimension 3 operator.  In the total rate this corresponds to a rescaling factor 
$1-\mu_\pi^2/2 m_b^2$ on the leading power result, 
which is nothing but the relativistic dilation factor of the lifetime of a moving quark and applies at any order in perturbation theory.
The relations for differential distributions and moments tend to be  more intricate, see 
\cite{Becher:2007tk}, especially in the presence of experimental cuts.

Recently Manohar has derived  elegant RI relations \cite{Manohar:2010sf} that apply 
directly at the  level of  
the structure functions $W_i$. They are also valid to all orders in perturbation theory and give the 
$\as^n$ coefficient of $\mu_\pi^2$ in terms of the  leading $\as^n$ coefficient and its derivatives:
\bea
W_1^{(\pi,n)}&=&- W_1^{(n)} +\frac23 W_2^{(n)}-2\hat q_0\, \frac{d\,W_1^{(n)}}{d\,\hat u}+\frac{\lambda}3  \frac{d^2 W_1^{(n)}}{d\,\hat u^2},\nn\\
W_2^{(\pi,n)}&=&\frac53 W_2^{(n)} -\frac{14}3\hat q_0\, \frac{d\,W_2^{(n)}}{d\,\hat u}+\frac{\lambda}3  \frac{d^2 W_2^{(n)}}{d\,\hat u^2},\label{RPI}\\
W_3^{(\pi,n)}&=& -\frac{10}3\hat q_0\, \frac{d\,W_3^{(n)}}{d\,\hat u}+\frac{\lambda}3  \frac{d^2 W_3^{(n)}}{d\,\hat u^2}.\nn
\eea
To verify these relations from Eq.~(\ref{NLO}) we need the first two derivatives of the plus distribution of Eq.~(\ref{plus1}).
They can be re-expressed in terms of the higher order plus distributions  
introduced in Eq.~(\ref{plusn}) and of delta functions:
\be
\left[\frac1{\hat u}\right]'_+= - \left[\frac1{\hat u^2}\right]_+ 
 +\delta(\hat u) - \delta'(\hat u),
\ee
\be
\left[\frac1{\hat u}\right]''_+= 2 \left[\frac1{\hat u^3}\right]_+ 
 -\delta(\hat u)+2\,\delta'(\hat u) -\frac32  \delta''(\hat u),
\ee
where we have neglected terms that do not contribute upon integration in the physical range
(\ref{physrange}). The coefficients $W_i^{(\pi,1)}$ obtained from Eq.~(\ref{NLO})
using the
RI relations agree with the results given in the previous Section.
Using Eqs.~(\ref{RPI}) one can also verify the relations between moments with and without cuts given in  \cite{Becher:2007tk}.

\section{Summary}
We have presented an analytic calculation of the $O(\as)$ corrections to the Wilson 
coefficient of the kinetic operator in inclusive $B\to X_c \ell\nu$ semileptonic decays, following and extending the method developed in Ref.~\cite{ewerth}.
We have confirmed the numerical results presented in \cite{Becher:2007tk} and reproduced the RI relations given by Manohar \cite{Manohar:2010sf}. We have provided several details
of the technique we have used; in particular, the Appendix contains all the master integrals.

The calculation represents the first  part of  a complete study of the 
perturbative corrections to the coefficients of the power suppressed dimension 5 and 6 operators, and has offered us  the opportunity to perform various checks.
Our technique is currently employed to compute the $O(\as \mu_G^2/m_b^2)$ 
corrections, as done already in the case of  $B\to X_s \gamma$ in \cite{ewerth}.

\section*{Acknowledgements}
We thank Lorenzo Magnea for useful discussions. PG is grateful to Thomas Mannel and his group at the University of Siegen for the warm hospitality in June 2012 and to the CERN Theory Unit  for hospitality during the final stage of this work.
Work partly supported by MIUR under contract 2010YJ2NYW$\_$006 
and by the DFG through the SFB/TR~9 grant ``Computational Particle Physics''.

\section*{Appendix}
\setcounter{equation}{0}
\renewcommand{\theequation}{A.\arabic{equation}}
\renewcommand{\thesubsection}{\Alph{subsection}}
\setcounter{section}{0}

In this Appendix we list the master integrals relevant to our calculation. Whenever appropriate, they are expanded in $\epsilon$ up to the order which is necessary in our calculation. As explained above,
we need both the imaginary and real parts of the master integrals.
However,  since their real parts always appear multiplied by $\delta(\hat{u})$ or 
its derivatives, the real parts and their derivatives
wrt $\hat u$  are only necessary at  $\hat u=0$, {\it i.e.}  with partonic kinematics. 
 We introduce
\be\label{master}
I_{(n_1,n_2,n_3)}= -i
\mu^{2\epsilon}
\int \frac{d^d k}{\pi^{d/2} } \frac1{(k^2+i\eta)^{n_1}[(k\!-\!p)^2\!-\!m_b^2+i\eta]^{n_2} [(k\!-\!p\!+\!q)^2-m_c^2+i\eta]^{n_3}} \, .
\ee
Let us also employ
$
\hat z= (p-q)^2/m_b^2 = \hat u +\rho$.
The massive tadpole integral is always real and is given by 
\be
I_{(0,1,0)}= m_b^2 \left(\frac{\mu^2}{m_b^2}\right)^{\ep} \left[\frac1{\epsilon} +1+ \epsilon \left(1+\frac{\pi^2}{12} \right)+ O(\ep^2)\right],
\ee
with $I_{(0,0,1)}$ given by the same expression with $m_b\leftrightarrow m_c$.
We have two distinct two-point functions. The first one develops an imaginary part for 
$(p-q)^2>m_c^2$:
\bea
I_{(1,0,1)}&=&  e^{\gamma_E \epsilon}\,\Gamma(\epsilon)
\left(\frac{\mu^2}{m_c^2}\right)^{\epsilon} \int_0^1 dx (1-x)^{-\epsilon} \left[1- \frac{\hat{z}}{\rho} \,x-i \eta \right]^{-\epsilon}
\\
&=&\left(\frac{\mu^2}{m_b^2}\right)^\epsilon \left\{ \zh^{\ep-1} \, \hat u^{1-2\ep} \left[ e^{i \pi \ep}  
\left( \frac1{\ep} +2 
\right)
-\frac1{2\ep} -1 
\right]\right.\nn\\
&&+ \frac{\zh+\rho}{2\zh} \left[ \frac1{\ep} +2-\frac{\hat u \ln\zh +2\rho\ln\rho}{\zh+\rho}
 \right] + O(\ep)
\Big\}.
\nn
\eea
The value and derivatives wrt $\hat u$ of $I_{(1,0,1)}$ at $\hat u=0$ can be readily obtained from the above integral representation:
 \be
   I_{(1,0,1)}\big  |_{\hat u=0}= -2 \rho  \frac{d I_{(1,0,1)}}{d \hat u}\big  |_{\hat u=0}=
  \left(\frac{\mu^2}{m_b^2}\right)^\epsilon \rho^{-\epsilon}\left[\frac1{\epsilon} +2 
  + O(\epsilon)\right],\nn
 \ee
 \be
  \frac{d^2 I_{(1,0,1)}}{d \hat u^2}\big  |_{\hat u=0}=\left(\frac{\mu^2}{m_b^2}\right)^\epsilon   \rho^{-2-\epsilon}\left[   \frac1{\epsilon} +1 
  + O(\epsilon)
 \right],\nn\ee
 \be
  \frac{d^3 I_{(1,0,1)}}{d \hat u^3}\big  |_{\hat u=0}=-3\left(\frac{\mu^2}{m_b^2}\right)^\epsilon   \rho^{-3-\epsilon}\left[   \frac1{\epsilon} +\frac12 
  + O(\ep)
 \right].\nn\ee

The second 2-point function is always real in the kinematic domain we are interested in. It can 
be directly expanded in $\ep$:
\bea
I_{(0,1,1)}&&=e^{\gamma_E \ep}\Gamma(\ep) \left(\frac{\mu^2}{m_b^2}\right)^{\ep}
\int_0^1 dx \left[x \rho + (1-x) (1-x \hat{q}^2) \right]^{-\ep}\\
&&= \left(\frac{\mu^2}{m_b^2}\right)^{\ep}\left[\frac1{\ep}+2  +\frac{1-\rho-\hat{q}^2}{2\hat{q}^2} \ln \rho + 
\frac{E_0 t_0}{\hat q^2} \ln \frac{1+t_0}{1-t_0}
+O(\ep)\right].\nn
\eea

The only three-point function is $I_{(1,1,1)}$;  we reduce it in the following way
\bea I_{(1,1,1)}&=& -e^{\gamma_E \ep}\frac{\ep\Gamma(\ep)}{m_b^2}  \left(\frac{\mu^2}{m_b^2}\right)^{\ep}\int_0^1 dx\, dy \,x^{-\ep} \left[x \,\chi-(1-y)(\hat u +i \eta)\right]^{-1-\ep} \\
&=&-\frac1{m_b^2} \left(\frac{\mu^2}{m_b^2}\right)^{\ep}  \int_0^1 \frac{dy}{\chi} \left[ \hat u^{-2\ep}  e^{2i\pi\ep}
\left(\frac{1}{2\ep}   +\frac12 \ln \frac{\chi}{(1-y)^2}  + O(\ep)\right)\nonumber  \right.\\
&&\left.
\hspace{4.5cm} -\frac1{2\ep}+\ln \bar{\chi}-\frac12 \ln \chi  + O(\ep)\right],\label{I111}\nn
\eea
where $\chi=\hat z +y (1-\hat{q}^2-\hat{z}) + \hat{q}^2 y^2$,
 $\bar \chi=\rho +y (1-\hat{q}^2-\rho) + \hat{q}^2 y^2$.
 
We need the real and imaginary parts  of $I_{(1,1,1)}$ in the physical semileptonic region, as well as its derivatives wrt $\hat u$ at $\hat u =0$. The latter can be  computed  from  the Feynman parameter integral 
or can  be expressed in terms of the master integrals at $\hat{u}=0$ 
by applying the derivative wrt $\hat{u}$ 
at the integrand level in (\ref{master}), see Eq.~(\ref{deriv}).
We  obtain, up to $O(\epsilon)$ terms,
\be 
I_{(1,1,1)}\Big |_{\hat u=0}= \frac1{2\,  m_b^2} \left(
\frac1{\epsilon} I_{1,0}- I_{4,0}\right) ,\nn
\ee
\be
\frac{d}{d\hat u}I_{(1,1,1)}\Big|_{\hat u=0}=
\frac1{\lambda_0 m_c^2}\left[(\rho - E_0)(
\frac{1}{\epsilon}  -2)  + \rho  (1 - E_0) (\frac{I_{1,0}}{\epsilon}-I_{4,0})+ E_0 \ln \rho
\right], \nonumber
\ee
\bea
\frac{d^2}{d\hat u^2}I_{(1,1,1)}\Big|_{\hat u=0}&=&\frac{2}{\lambda_0^2 m_b^2}
\left\{
\frac{6\alpha-\beta/\rho^2}{4\,\epsilon} + \left(\frac{\alpha^2}2+\hat q^2\right)\left(\frac{I_{1,0}}{\epsilon} -I_{4,0}\right) + \frac{\beta}{4\rho^2} \left(\ln \rho+2\right)
  \right. \nn \\&&
\left.  
-\frac72\alpha -\frac12 +\frac{1-\hat q^2 (1+2\,\rho I_{1,0})}{2\rho} \right\},\nn
\eea
where $\beta =
-8E_0^2(E_0+\rho)+4\rho(5E_0-\rho)$, 
$\alpha =2(1-E_0)$, 
 and
\be 
I_1=  \int_0^1 dy \frac{1}{\chi(y)}= \frac{\ln\frac{1+t}{1-t}}{\sqrt{\lambda} } ,\quad\qquad\qquad I_{1,0}=\int_0^1 dy \frac{1}{\bar\chi(y)}= \frac{\ln\frac{1+t_0}{1-t_0}}{\sqrt{\lambda_0} },
\label{I1}
 \ee
 \bea 
I_{4,0}&=&\int_0^1 dy \frac{\ln \bar\chi(y)}{\bar\chi(y)}
= \frac1{t_0 \, E_0} \Big[ {\rm Li}_2\left(a_1
\right)\!-\!{\rm Li}_2\left(a_2\right) \nn\\&&\left.
+\ln\frac{1\!+\!t_0}{1\!-\!t_0} \ln E_0 (1\!-\!t_0)^{\frac14} (1\!+\!t_0)^{\frac34} +\ln E_0(1\!+\!t_0) \ln\frac{1\!-\!E_0(1\!-\!t_0)}{1\!-\!E_0(1\!+\!t_0)}
\right]\label{I40}
 \eea
 and
 \bea &&
 a_1= \frac{2t_0/(1+t_0)}{1-E_0(1-t_0)},\quad\quad a_2=\frac{2t_0\, E_0}{1-E_0(1-t_0)},\nn
 \eea
For the imaginary part we also need the integral 
\be 
I_{\chi}=\int_0^1 dy\, \frac{\ln[\chi(y)/(1-y)^2]}{\chi(y)} \nn
\ee
and its first derivatives at $\hat u=0$:
\be
I_{\chi}\Big |_{\hat u=0}=I_{4,0}-2 I_{2,0},\nn
\ee
 \be
 \frac{d I_\chi}{d\hat u}\Big|_{\hat u=0}=  \frac{2(1-E_0)}{\lambda_0}\left(   I_{4,0}-2I_{1,0}-2I_{2,0}\right)   +2 \frac{(\rho-E_0)}{\lambda_0} \frac{\ln \rho}{\rho},\nn
 \ee
 
  \be
 \frac{d^2 I_\chi}{d\hat u^2}\Big|_{\hat u=0}=  
  \frac{\left(3\hat q^2 +2 E_0^2 t_0^2\right)\left(   I_{4,0}-3I_{1,0}-2I_{2,0}\right)
  + \hat q^2 (I_{1,0}-E_0/\rho)}{\lambda_0^2/4} -\frac{\beta-6 \alpha\rho^2}{\rho^2\,\lambda_0^2}
\left(\ln \rho -1\right) ,
 \nn
 \ee
where
\be 
I_{2,0}=\int_0^1 dy \frac{\ln (1-y)}{\bar\chi(y)}=  \frac{{\rm Li}_2(1-E_0(1+t_0))-{\rm Li}_2(1-E_0(1-t_0))}{\sqrt{\lambda_0} }.
\label{I20}
 \ee


\begin{thebibliography}{99}


\bibitem{Bigi:1992su}
  I.~I.~Y.~Bigi, N.~G.~Uraltsev and A.~I.~Vainshtein,
  Phys.\ Lett.\ B {\bf 293} (1992) 430 [Erratum-ibid.\  B {\bf 297} (1993) 477]
  [arXiv:hep-ph/9207214];
  I.~I.~Y.~Bigi, M.~A.~Shifman, N.~G.~Uraltsev and A.~I.~Vainshtein,
  Phys.\ Rev.\ Lett.\ {\bf 71} (1993) 496 [arXiv:hep-ph/9304225].

\bibitem{Blok:1993va}
  B.~Blok, L.~Koyrakh, M.~A.~Shifman and A.~I.~Vainshtein,
  Phys.\ Rev.\  D {\bf 49} (1994) 3356 [Erratum-ibid.\ D {\bf 50} (1994) 3572]
  [arXiv:hep-ph/9307247];
  A.~V.~Manohar and M.~B.~Wise, Phys.\ Rev.\ D {\bf 49} (1994) 1310 [arXiv:hep-ph/9308246].

\bibitem{fits} C.~W.~Bauer, Z.~Ligeti, M.~Luke, A.~V.~Manohar and M.~Trott,
  Phys.\ Rev.\ D {\bf 70} (2004) 094017
  [hep-ph/0408002];
  O.~Buchmuller, H.~Flacher,
  Phys.\ Rev.\  {\bf D73 } (2006)  073008.
  [hep-ph/0507253].
  
\bibitem{ckm10}
  P.~Gambino, C.~Schwanda,
    [arXiv:1102.0210 [hep-ex]].
      
  \bibitem{hfag}
  Y.~Amhis {\it et al.}  [Heavy Flavor Averaging Group Collaboration],
  arXiv:1207.1158 [hep-ex], see also {\tt http://www.slac.stanford.edu/xorg/hfag/} .

\bibitem{Antonelli:2009ws}
  M.~Antonelli {\it et al.},
  Phys.\ Rept.\  {\bf 494} (2010) 197
  [arXiv:0907.5386 [hep-ph]].


\bibitem{bsgamma}
M.~Misiak,  {\it et al.},
  Phys.\ Rev.\ Lett.\  {\bf 98} (2007) 022002
  [hep-ph/0609232];
M.~Misiak and M.~Steinhauser,
  Nucl.\ Phys.\ B {\bf 764}, 62 (2007)
  [hep-ph/0609241].
\bibitem{review}
T.~Hurth and M.~Nakao,
  Ann.\ Rev.\ Nucl.\ Part.\ Sci.\  {\bf 60} (2010) 645
  [arXiv:1005.1224 [hep-ph]].

\bibitem{gg}
  P.~Gambino, P.~Giordano,
  Phys.\ Lett.\  {\bf B669 } (2008)  69-73.
  [arXiv:0805.0271 [hep-ph]].
  
\bibitem{1mb3}
M.~Gremm and A.~Kapustin,
  Phys.\ Rev.\ D {\bf 55} (1997) 6924
  [hep-ph/9603448].

\bibitem{Mannel:2010wj}
  T.~Mannel, S.~Turczyk and N.~Uraltsev,
  JHEP {\bf 1011}, 109 (2010)
  [arXiv:1009.4622 [hep-ph]].
  
  \bibitem{Gambino:2012rd}
    P.~Gambino, T.~Mannel and N.~Uraltsev,
  JHEP {\bf 1210} (2012) 169
  [arXiv:1206.2296 [hep-ph]].

\bibitem{pert}
M.~Jezabek and J.~H.~Kuhn,
Nucl.\ Phys.\ B {\bf 314} (1989) 1;
A.~Czarnecki, M.~Jezabek and J.~H.~Kuhn,
Acta Phys.\ Polon.\ B {\bf 20} (1989) 961;
A.~Czarnecki and M.~Jezabek,
Nucl.\ Phys.\ B {\bf 427} (1994) 3
[arXiv:hep-ph/9402326];
M.~Gremm and I.~Stewart,
{\it Phys.\ Rev.}\ {\bf D55} (1997) 1226;
 A.~F.~Falk, M.~E.~Luke,
  Phys.\ Rev.\  {\bf D57 } (1998)  424
  [hep-ph/9708327];
  A.~F.~Falk, M.~E.~Luke, M.~J.~Savage,
  Phys.\ Rev.\  {\bf D53 } (1996)  2491
  [hep-ph/9507284];
M.~Trott,
  Phys.\ Rev.\  D {\bf 70} (2004) 073003
  [arXiv:hep-ph/0402120].


 \bibitem{Aquila}
  V.~Aquila, P.~Gambino, G.~Ridolfi and N.~Uraltsev,
  Nucl.\ Phys.\  B {\bf 719} (2005) 77
  [arXiv:hep-ph/0503083].


\bibitem{melnikov}
  K.~Melnikov,
  Phys.\ Lett.\  B {\bf 666} (2008) 336
  [arXiv:0803.0951 [hep-ph]].
  
  \bibitem{czarnecki-pak}
  A.~Pak and A.~Czarnecki,
  Phys.\ Rev.\ Lett.\  {\bf 100} (2008) 241807
  [arXiv:0803.0960 [hep-ph]];  Phys.\ Rev.\  {\bf D78 } (2008)  114015.
  [arXiv:0808.3509 [hep-ph]].

  \bibitem{melnikov2}
 S.~Biswas and K.~Melnikov,
  JHEP {\bf 1002} (2010) 089
  [arXiv:0911.4142 [hep-ph]].

  \bibitem{Gambino:2011cq}
  P.~Gambino,
  JHEP {\bf 1109} (2011) 055
  [arXiv:1107.3100 [hep-ph]].
  
  \bibitem{Becher:2007tk}
  T.~Becher, H.~Boos and E.~Lunghi, JHEP {\bf 0712} (2007) 062 [arXiv:0708.0855].
\bibitem{RPI}
  M.~E.~Luke and A.~V.~Manohar,
  Phys.\ Lett.\ B {\bf 286} (1992) 348
  [hep-ph/9205228].

  \bibitem{Manohar:2000dt}
  A.~V.~Manohar and M.~B.~Wise, Camb.\ Monogr.\ Part.\ Phys.\ Nucl.\ Phys.\ Cosmol.\ {\bf 10} (2000) 1.

\bibitem{Manohar:2010sf}
  A.~V.~Manohar,
  Phys.\ Rev.\ D {\bf 82} (2010) 014009
  [arXiv:1005.1952 [hep-ph]].

  \bibitem{ewerth}
 T.~Ewerth, P.~Gambino and S.~Nandi,
  Nucl.\ Phys.\  B {\bf 830} (2010) 278
  [arXiv:0911.2175 [hep-ph]].


\bibitem{IBP}
  F.~V.~Tkachov, Phys.\ Lett.\ B {\bf 100}, 65 (1981);
  K.~G.~Chetyrkin and F.~V.~Tkachov, Nucl.\ Phys.\ B {\bf 192}, 159 (1981);
  S.~Laporta, Int.\ J.\ Mod.\ Phys.\ A {\bf 15}, 5087 (2000) [arXiv:hep-ph/0102033].


\bibitem{larin} 
 S.~A.~Larin,
  Phys.\ Lett.\ B {\bf 303} (1993) 113
  [hep-ph/9302240].
  
 
\end{thebibliography}
\end{document}